\documentclass[aps,twocolumn,prl,superscriptaddress,nofootinbib,showpacs,letterpaper]{revtex4}

\usepackage{amsmath,amssymb}
\usepackage[dvips]{graphicx}
\usepackage{longtable}
\usepackage[usenames]{color}
\usepackage[normalem]{ulem}
\usepackage{bm}

\begin{document}

\newcommand{\comment}[1]{}
\newcommand{\E}{\mathrm{E}}
\newcommand{\Var}{\mathrm{Var}}
\newcommand{\bra}[1]{\langle #1|}
\newcommand{\ket}[1]{|#1\rangle}
\newcommand{\braket}[2]{\langle #1|#2 \rangle}
\newcommand{\mean}[2]{\langle #1 #2 \rangle}
\newcommand{\be}{\begin{equation}}
\newcommand{\ee}{\end{equation}}
\newcommand{\ba}{\begin{eqnarray}}
\newcommand{\ea}{\end{eqnarray}}
\newcommand{\SD}[1]{{\color{magenta}#1}}
\newcommand{\rem}[1]{{\sout{#1}}}
\newcommand{\alert}[1]{\textbf{\color{red} \uwave{#1}}}
\newcommand{\Y}[1]{\textcolor{blue}{#1}}
\newcommand{\R}[1]{\textcolor{red}{#1}}
\newcommand{\B}[1]{\textcolor{blue}{#1}}
\newcommand{\C}[1]{\textcolor{cyan}{#1}}
\newcommand{\db}{\color{darkblue}}
\newcommand{\intinfty}{\int_{-\infty}^{\infty}\!}
\newcommand{\Tr}{\mathop{\rm Tr}\nolimits}
\newcommand{\const}{\mathop{\rm const}\nolimits}

\title{Preparing a mechanical oscillator in non-Gaussian quantum states}

\author{Farid Khalili}
\affiliation{Physics Faculty, Moscow State University, Moscow
119991, Russia}
\author{Stefan Danilishin}
\affiliation{Physics Faculty, Moscow State University, Moscow
119991, Russia}
\author{Haixing Miao}
\affiliation{School of Physics, University of Western Australia,
WA 6009, Australia}
\author{Helge M\"uller-Ebhardt}
\affiliation{Max-Planck Institut f\"ur Gravitationsphysik
(Albert-Einstein-Institut) and Leibniz Universit\"at Hannover,
Callinstr. 38, 30167 Hannover, Germany}
\author{Huan Yang}
\affiliation{Theoretical Astrophysics 130-33, California Institute
of Technology, Pasadena, CA 91125, USA}
\author{Yanbei Chen}
\affiliation{Theoretical Astrophysics 130-33, California Institute
of Technology, Pasadena, CA 91125, USA}

\begin{abstract}
We propose a protocol for coherently transferring non-Gaussian
quantum states from optical field to a mechanical oscillator.
The open quantum dynamics and continuous-measurement process, which
can not be treated by the stochastic-master-equation formalism, are
studied by a new path-integral-based approach. We obtain an elegant
relation between the quantum state of the mechanical oscillator
and that of the optical field, which is valid for general linear
quantum dynamics. We demonstrate the experimental feasibility of
such protocol by considering the cases of both large-scale
gravitational-wave detectors and small-scale cavity-assisted
optomechanical devices.
\end{abstract}
\maketitle

{\it Introduction.}---It is becoming experimentally possible to
prepare a macroscopic mechanical oscillator near its quantum ground
state by either active feedback or passive cooling in
optomechanical devices~\cite{Marquardt}. This activity has been
motivated by (i) the necessity to increase the sensitivity of
high-precision measurements with mechanical test bodies up to
and beyond the {\it Standard Quantum Limit} (SQL)~\cite{92BookBrKh},
and (ii) the test and interpretation of quantum theory, when
macroscopic degrees of freedom are involved. However, for unequivocal
evidences of quantum behavior, {\it merely} achieving quantum
ground state, or preparing coherent/squeezed states, or
overcoming the SQL is {\it insufficient}: In these situations,
the oscillator initially occupies a Gaussian state and remains
Gaussian, and therefore its Wigner function is positive and can
always be interpreted in terms of a classical probability.
{\it A true demonstration of the quantum behavior requires non-Gaussian
quantum states or nonlinear measurements} \cite{Bell1987, Braunstein2005}. A natural approach is
to create nonlinear coupling between a mechanical oscillator and
external degrees of freedom, e.g., probing mechanical energy
\cite{Thompson, Santamore, Martin, Miao}, coupling the oscillator to
a qubit \cite{Jacobs,Clerk,LaHaye} or (low) cavity photon number
\cite{Mancini, Bose, Marshall}. For optomechanical devices,
this generally requires zero-point uncertainty of the oscillator
displacement $x_q$ to be comparable to the cavity linear dynamical
range which is characterized by the optical wavelength $\lambda$
divided by the finesse $\cal F$, i.e.,
\be\label{cond1}
\lambda/({\cal F} x_q)\lesssim 1.
\ee
Since $\lambda\sim 10^{-6}$m and ${\cal F}\lesssim10^6$, we have
$x_q\gtrsim 10^{-12}\,{\rm m}$, which is several orders of magnitude
above the current technology ability.
\begin{figure}
\includegraphics[width=0.48\textwidth, bb=0 0 270 125,clip]{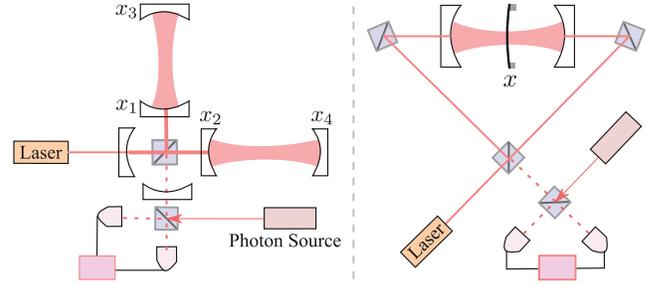}
\caption{(Color online) Possible schemes for preparing non-Gaussian quantum
states of mechanical oscillators. The left is a Michelson interferometer,
similar to an advanced gravitational-wave detector with kg-scale suspended
test masses \cite{LIGO, AdvLIGOsite}. The right panel shows a small
coupled-cavity scheme with a ng-scale membrane inside a high-finesse cavity
\cite{Thompson}. In both cases, a non-Gaussian optical state (a photon pulse)
is injected into the dark port of the interferometer (local oscillator light
for homodyne detection is not shown).
\label{config}}
\end{figure}

In this article, we propose a protocol for preparations of non-Gaussian
quantum states which {\it does not require} nonlinear optomechanical
coupling. The idea is to inject a non-Gaussian optical state, e.g.,
a single-photon pulse created by cavity QED \cite{Kimble1, Kimble2, Walther},
into the optomechanical devices. Possible configurations are shown schematically
in Fig. \ref{config}. The radiation pressure induced by the photon pulse
is coherently amplified by the classical pumping at the bright port, and
the qualitative requirement for preparing a non-Gaussian state is
\be\label{cond2}
\lambda /({\cal F}\,x_q)\lesssim \sqrt{N_\gamma}.
\ee
Here $N_\gamma=I_0\,\tau/(\hbar \omega_0)$ ($I_0$ the pumping laser power and
$\omega_0$ the frequency) is the number of pumping photons within
the duration $\tau$ of the single-photon pulse, and we gain a significant
factor of $\sqrt{N_\gamma}$ compared with Eq.~\eqref{cond1}, which makes
it experimentally achievable. This radiation-pressure-mediated optomechanical coupling is similar to what
was considered in Refs. \cite{92BookBrKh,Zhang2003, Mancini2, Romero}.
However, there are significant differences: (i) This protocol includes both finite interaction time
and photon shape, in which case neither the rotating-wave approximation
\cite{Zhang2003} nor the three-mode
approach \cite{Mancini2} applies; (ii)
To better model an actual experiment, we consider a continuous measurement process rather than a
single measurement at some given instant as assumed in Ref. \cite{Romero}. This
takes into account all the information of the oscillator motion that is distributed in the output field, and thus
allows us to prepare a nearly {\it pure} non-Gaussian quantum state of the oscillator;
(iii) There are non-trivial quantum correlations at different times (non-Markovianity)
due to the finite-duration photon pulse, which cannot be treated by
the conventional {\it stochastic-master-equation} (SME) approach \cite{Hopkins, Gardiner, Milburn, Doherty1, Doherty2}.
Here we develop a path-integral-based approach, and it applies to general
linear quantum dynamics and continuous measurement process.

{\it A simple case.}---To illustrate the non-Gaussian state-preparation
procedure, we first make an order-of-magnitude estimate of experimental
requirements by considering a simple case where the cavity decay is
much faster than all other time scales and the oscillator can be
approximated as a free mass. The corresponding input-output relations,
in the Heisenberg picture, simply read:
\begin{align}
   &\dot{\hat x}(t) = {\hat p}(t)/m \,,\quad\;\;
   \dot{\hat p}(t) = \alpha\,\hat a_1(t) + \hat F_{\rm th}(t) \,,\label{1} \\
   &\hat b_1(t) = \hat a_1(t)\,,\quad\quad
   \hat b_2(t) =\hat a_2(t) + ({\alpha}/{\hbar})\hat x(t) \,. \label{2}
\end{align}
Here $\hat x$ and $\hat p$ are position and momentum; the coupling
constant {$\alpha\equiv 8\sqrt{2}({\cal F}/\lambda) \sqrt{\hbar I_0/\omega_0}$;
$\hat a_{1,2}$ and $\hat b_{1,2}$ are input and output optical
amplitude and phase quadratures,
with $[\hat a_1(t),\hat a_2(t')]=[\hat b_1(t),\hat b_2(t')] = i\,\delta(t-t')$;
$\alpha\, \hat a_1$ is the back-action noise; $\hat F_{\rm th}$ is the
force thermal noise.

Suppose at $t=-\tau$ the oscillator was prepared in some initial Gaussian
state $\ket{\psi_m}=\intinfty\psi_m(x)\ket{x}\,dx$ (the procedure is detailed
in Ref. \cite{state_pre}).
Subsequently, a photon pulse is injected into the dark port of the interferometer
and starts to interact with the oscillator. During this interaction, phase
quadrature $\hat b_2(t)$ is continuously measured by a homodyne detection,
until the photon pulse ends at $t=0$. If photon pulse (i.e., $\tau$) is short such that
oscillator position almost does not change, we obtain:
\begin{align}\label{3}
&\hat{X}(0)=\hat{X}(-\tau),\quad\hat{P}(0) = \hat{P}(-\tau) + \kappa\,\hat{A}_1
+ \hat P_{\rm th} \,, \\\label{4}
&\hat{B}_1 =\hat{A}_1,\quad\quad\quad\;\;\, \hat{B}_2 = \hat{A}_2 + \kappa\,\hat{X}(0) \,.
\end{align}
We have normalized the oscillator position and momentum by their zero-point
uncertainties: $\hat X\equiv \hat x/ x_q$ [$x_q\equiv\sqrt{\hbar/(2m\omega_m)}$]
and $\hat P\equiv \hat p/p_q$ [$p_q\equiv \sqrt{\hbar m\omega_m/2}$]; $\hat{A}_j =
\sqrt{1/\tau}\int_{-\tau}^0dt\,\hat{a}_j(t)\,(j=1,2)$ which has an uncertainty of unity
(i.e., $\Delta \hat A_j$=1); $\hat{B}_j = \sqrt{1/\tau}\int_{-\tau}^0dt\,\hat{b}_j(t)$;
$\hat{P}_{\rm th} = \int_{-\tau}^0 dt\,\hat{F}_{\rm th}(t)/p_q$; $
\kappa \equiv {\alpha\sqrt{\tau}}/{\hbar}=8\sqrt{2}\sqrt{N_{\gamma}}{{\cal F}\,x_q}/{{\lambda}}$.

Eqs.~\eqref{3} and \eqref{4} describe the joint evolution of the oscillator, the
optical field and heat bath in the Heisenberg picture (with $\hat B_j$ viewed
as the evolved versions of $\hat A_j$). They transform back into an evolution
operator of $\hat{U}= \exp[i(\kappa\hat{A}_1\hat{X} + \hat{P}_{\rm th}\hat{X})]$
in the Schr\"odinger picture. The corresponding density matrix of the system
at $t=0$ is given by $ \hat \rho=\hat{U}\ket{\psi_o}\ket{\psi_m}\hat{\rho}_{\rm th}\bra{\psi_m}\bra{\psi_o}\hat{U}^\dagger$, where $\ket{\psi_o} = \intinfty\psi_o(A_2)\ket{A_2}\,dA_2$ is the initial
non-Gaussian optical state, and $\hat{\rho}_{\rm th}$
describes the heat bath associated with $\hat F_{\rm th}$. Given homodyne
detection of $\hat B_2$ with a precise result $y$, the oscillator is projected
into the following conditional state: $\hat{\rho}_m(y) = \Tr_{\rm th}\left[ \bra{y}\hat{U}\ket{\psi_o}\ket{\psi_m}\hat{\rho}_{\rm th}\bra{\psi_m}\bra{\psi_o}\hat{U}^\dagger \ket{y} \right]$.
In the ideal case of negligible thermal noise, the conditional
wave function $\psi_m^c(x)$ of the mechanical oscillator is simply
\begin{equation}
\psi_m^c(x)=\psi_o(y-\kappa x)\psi_m(x) \,
\end{equation}
---{\it the optical state is mapped onto the mechanical oscillator} as
illustrated in Fig. \ref{Int}. A complete mapping occurs when
$\psi_m(x)\approx\const$, and this requires the momentum fluctuation
due to optomechanical coupling be larger than the initial one, namely,
 $\kappa >1$ or equivalently
\be\label{8}
{\lambda}/({{{\cal F}\,x_q}})<8\sqrt{2}{\sqrt{N_{\gamma}}},
\ee
which justifies Eq.\,\eqref{cond2}.

\begin{figure}
\includegraphics[width=0.4\textwidth, bb=0 0 422 229,clip]{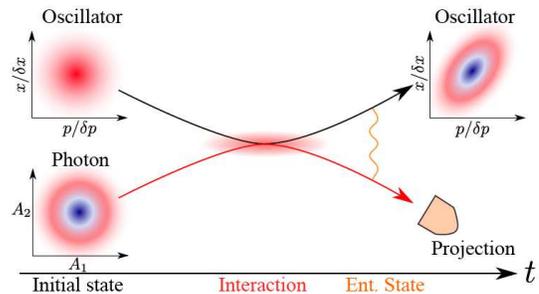}
\caption{(Color online) A schematic of the non-Gaussian state-preparation
process. The interaction entangles the oscillator state and the optical state
(depicted by their Wigner functions). Subsequent measurements of the optical
fields disentangle the system and projects the oscillator into a non-Gaussian
conditional quantum state.
\label{Int}}
\end{figure}

When thermal noise is considered, non-Gaussianity can still remain, as long as thermal noise induces a
smaller momentum fluctuation than the optomechanical interaction. This condition, in the high-temperature limit
--- $\langle \hat F_{\rm th}(t)\hat F_{\rm th}(t')\rangle=4m\gamma_m k_B T\delta (t-t') $,
reads
\be\label{9}
{\lambda}/({{\cal F}\,x_q})\sqrt{n_{\rm th}/Q_m}\sqrt{\omega_m \tau}< 8\sqrt{2}\sqrt{N_{\gamma}}
\ee
with $Q_m\equiv \omega_m/\gamma_m$ the mechanical quality factor and
$n_{\rm th}\equiv k_B T/(\hbar\,\omega_m)$ the thermal occupation number.
{\it These two conditions set the benchmarks for a successful non-Gaussian
state-preparation experiment.} They can be satisfied with experimentally
feasible specifications as shown in Table \ref{tab1},
in which the first row is similar to the case of large-scale gravitational-wave detectors \cite{AdvLIGOsite}
\begin{table}[!h]
\caption{Possible experimental specifications}\label{tab1}
\begin{tabular}{l|ccccccc}
&$\lambda$& ${\cal F}$ & $m$ & $\omega_m/2\pi$ &  $Q_m$ & $T$ & $\tau$\\
 \hline
 large scale & $1\mu{\rm m}$ & $6000$ & 4kg & 1Hz & $10^8$ & 300K & 1ms\\
 small scale & $1\mu{\rm m}$ & $10^4$ & 1ng & $10^5$Hz & $10^7$ & 4K & 0.01ms
\end{tabular}
\end{table}
and the second row is for small-scale optomechanical devices (e.g., the
one in Ref. \cite{Thompson}). These qualitative results will be justified
by a rigorous treatment below.

{\it General formalism.}---In general, the optomechanical interaction strength is
finite and the oscillator has non-negligible displacement during the
interaction, the cavity bandwidth can be comparable to the mechanical
frequency, and thermal noises can be non-Markovian. All these factors
obstruct finding a finite set of variables similar to $(\hat X, \hat P, \hat A_1, \hat A_2)$
that satisfy a closed set of equations [cf. Eqs.~\eqref{3} and
\eqref{4}]. It is therefore hard to determine, {\it a priori}, the finite
number of observables that one has to measure to project the oscillator into
a desired conditional state.

To address these issues, we adopt the Heisenberg picture starting from $t=-\infty$, and write down
the initial density matrix as  $\hat \rho_{in} =\hat \rho_m(-\infty)
\otimes\hat \rho_o\otimes\hat \rho_{\rm th}.$ Details of $\hat\rho_m(-\infty)$
for the oscillator and whether the initial state is truly a direct product, do not
matter, because the system is stable, and the initial position and momentum will decay away
after several mechanical relaxation
times. For the optical state, we consider an arbitrary spatial mode
given by $f(x/c)$, whose annihilation operator is
\begin{equation}
\hat \Gamma\equiv \textstyle \int_{-\infty}^0 dt{f(t)}[\hat a_1(t)+i\hat a_2(t)]/{\sqrt{2}}.
\end{equation}
A general state of this mode can be written in the P-representation as
$\hat \rho_o=\int d\bm\zeta\, P(\bm \zeta)|\zeta\rangle \langle \zeta|$,
where vector $\bm \zeta\equiv(\Re[\zeta],\Im[\zeta])$ and
$|\zeta\rangle\equiv\exp[\zeta\,\hat \Gamma^{\dag}-\zeta^*\hat\Gamma]|0\rangle$.

A continuous measurement of the output optical quadrature
$\hat y(t)\equiv \cos\theta \,\hat b_1(t)+\sin\theta \,\hat b_2(t)$
for  $t \in (-\infty,0]$, {\it projects the entire system into a
conditional state:}
\be\label{rhom}
\hat \rho_c[y(t)]={\hat {\cal P}_y\,\hat \rho_{in} \hat {\cal P}_y}
/{\mathrm{Tr}[\hat {\cal P}_y\,\hat \rho_{in} \hat {\cal P}_y]}.
\ee
The operator $\hat {\cal P}_y$ projects the output field into the
subspace where $\hat y(t)$ agrees exactly with the measured results
$y(t)$. To simplify output correlations at different times, we can
{\it causally whiten} $\hat y(t)$ into $\hat z(t)$ such that
$\langle \hat z(t)\hat z(t')\rangle=\delta(t-t')$, as detailed in
Ref. \cite{state_pre}. Since the output quadratures at different
times also commute, i.e., $[\hat z(t), \hat z(t')]=0$, the projection
$\hat {\cal P}_y$ can then be expressed as the product of
Dirac-$\delta$ functions that project each $\hat z(t)$ into its
measured value $z(t)$:
\begin{align}\nonumber
\hat {\cal P}_y =\hat {\cal P}_z&=\prod_{-\infty<t<0}\delta[\hat z(t)-z(t)]\\&=\int{\cal D}[\xi]\exp\left\{i\mbox{$\int_{-\infty}^0$}dt\,\xi(t)[\hat z(t)-z(t)] \right\}.\label{12}
\end{align}
with $\int {\cal D}[\xi]$ denoting the path integral.
This allows us to take the entire measurement history
for $z$ (or equivalently $y$) and project into the corresponding
subspace in a single step, instead of having to successively project
output-field degrees of freedom continuously at each time step as
in the case of SME approach, thereby allowing a non-Markvonian input field.

The generating function for the oscillator state is then
\be
{\cal J}[\bm \alpha; z(t)]\equiv {\rm Tr}\left[e^{i \,\bm \alpha\,
\hat{\bm x}_0'}{\hat \rho_c[z(t)]}\right],
\ee
where ${\bm \alpha}\equiv(\alpha_x, \alpha_p)$, $\hat{\bm x}_0\equiv
(\hat x(0),\hat p(0))$, and superscript $'$ denotes transpose. From
Eqs. \eqref{rhom} and \eqref{12}, we have
\be
{\cal J}={\int d\bm \zeta\,{P(\bm \zeta)}\int{\cal D}[\xi]\,
e^{i[\zeta^*\hat\Gamma-\zeta\,\hat \Gamma^{\dag},\,\hat B]}\langle 0|e^{i\hat B} |0\rangle}
\label{J}
\ee
with $\hat B\equiv  \bm\alpha \,\hat{\bm x}_0' +\int_{-\infty}^0 dt\,\xi(t)[\hat z(t)-z(t)]$.
This can be evaluated by decomposing
$
\hat{\bm {x}}_0 \equiv \hat{\bm R}+\int_{-\infty}^0 dt\, \bm K(-t)\hat z(t)$
where $\bm K \equiv (K_x,K_p)$ are causal Wiener filters,
$\bm {K}(-t) = \langle 0 |\hat z(t)   {\bm {\hat x}_0}|0\rangle$ and
$\hat{\bm R} \equiv (\hat R_x ,\hat R_p)$ are parts of displacement and
momentum uncorrelated with the output: $\bra{0}\hat R_{x,p} \hat z\ket{0}=0$.
Completing path integral, we obtain
{\begin{eqnarray}
{\cal J}\!=\!\!\!\int d\bm \zeta e^
{-[{\bm\alpha{\mathbb V}_c\bm \alpha' +\|z-2\,\bm \zeta\bm L' \|^2 }]/2+ i\,
\bm\alpha\,(\zeta^*\gamma'+\zeta\gamma^\dagger +\bm x_c')
}P(\bm \zeta). \;
\end{eqnarray}
Here $\|a\|^2\equiv\int_{-\infty}^0a(t) a^*(t) dt $ and we have
defined vectors $\bm{\gamma}\equiv [\hat\Gamma, \hat{ \bm R}]$ and
$\bm L\equiv (\Re[L], \Im[L])$ with $L(t)\equiv [\hat\Gamma, \hat z(t)]$,
which characterize the extent of photon mode influence on the
fluctuations of $\hat x(0)$ and $\hat p(0)$, and output field $\hat z$;
quantities ${\mathbb V}_c\equiv \langle 0|\bm {\hat R}^T\bm {\hat R}|0\rangle$
and  $\bm x_c\equiv(x_c, p_c)=\int_{-\infty}^0 dt \bm K(-t)z(t)$ are
the conditional covariance matrix and means of $\hat x(0)$ and $\hat p(0)$
when the optical state is vacuum. The resulting {\it conditional} Wigner function reads
\begin{equation}
W[{\bm x}; z(t)] =\int d\bm\zeta  e^{-[{{\bm \chi}{\mathbb V}_c^{-1}{\bm \chi}'
+ \|z-2\,\bm\zeta\bm L'\|^2}]/{2}}P(\bm\zeta)\,\label{W}
\end{equation}
with $\bm \chi\equiv {\bm x} - {\bm x}_c-\zeta^* {\bm \gamma}-\zeta {\bm \gamma}^*$.
{\it This formula directly relates the injected optical state to the state of
the mechanical oscillator.} In deriving it, we only use the linearity of
quantum dynamics rather than specific equations of motion. For cavity-assisted
optomechanical system, one can obtain $\bm\gamma$, ${\mathbb V}_c$, $\bm K$
and $L$ from input-output relations in Refs.~\cite{Marquardt2, Rae, Genes}
by using formalism developed in Ref.~\cite{state_pre}.}

{\it Single-photon case.}---As an example, we consider the simplest case of a single-photon injection,
with $\hat \rho_o=|1\rangle \langle 1|$ and $P(\bm\zeta) =e^{|\zeta|^2}
\partial^2 \delta^{(2)}(\zeta)/\partial \zeta \partial \zeta^*$.
From Eqs. \eqref{W}, it gives
\begin{eqnarray}\nonumber
W[{ \bm x}; z(t)]&=&\frac{1-\bm \gamma \mathbb V_c^{-1}\bm \gamma^{\dag}-\| L \|^2
+|\bm \gamma V_c^{-1}\delta{\bm x}'+Z|^2}{1-\| L\|^2+|Z|^2}\nonumber \\&&
\frac{1}{2\pi\sqrt{\det {\mathbb V}_c}}\exp\left[-\frac{1}{2}\delta {\bm x}
{\mathbb V}_c^{-1}\delta {\bm x}'\right]\label{Wig}
\end{eqnarray}
where $\delta \bm x\equiv {\bm x} - {\bm x}_c$ and $Z \equiv\int_{-\infty}^0dt\,z(t)L(t)$.
This Wigner function depends on the measurement result $z(t)$, $t\in(-\infty,0]$ through
four quantities, the two components of ${\bm x}_c$ (through $\delta \bm{x}$) and
the real and imaginary parts of $Z$: $Z$ determines the shape of  $W$, and ${\bm x}_c$
describes the translation of $W$. The random vector $\bm Z= (\Re[Z], \Im[Z])$ has a
two-dimensional probability density of \be w[\bm Z]=\frac{1-\| L\|^2 +\bm Z\bm Z'}{2\pi
\sqrt{\det {\mathbb V}_L}} \exp[-\bm Z {\mathbb V}^{-1}_L \bm Z'/2], \ee where matrix
${\mathbb V}_L\equiv \int_{-\infty}^0 dt \bm{L}'\bm{L}$.

The pre-factor in the Wigner function [cf. Eq.\,\eqref{Wig}] is a second-order
polynomial in $\bm x$, which resembles that of a single-photon. For
strong non-Gaussianity, significant $\bm \gamma$ and $\|L\|^2$ (making $\bm \gamma$
terms in the pre-factor to prevail) are essential --- these physically
correspond to requiring that the photon mode must influence the fluctuation
of $\hat x$ and $\hat p$, as well as $\hat z$ strongly. It in turn requires the photon
coherence time to be comparable to the {\it measurement time scale} characterized
by $\bm K$. It is possible for small-scale optomechanical devices with
high-frequency mechanical oscillators. The corresponding photon can be generated
by a cavity QED scheme \cite{Kimble1, Kimble2, Walther}. While for large-scale
gravitational-wave detectors, the time scale is $\sim 10$ ms and it is challenging to create
photons with comparable coherent length. However, developments of low-frequency
squeezing source \cite{Corbitt2006} will eventually solve this issue.

\begin{figure}
\includegraphics[width=0.48\textwidth, bb=0 0 745 250,clip]{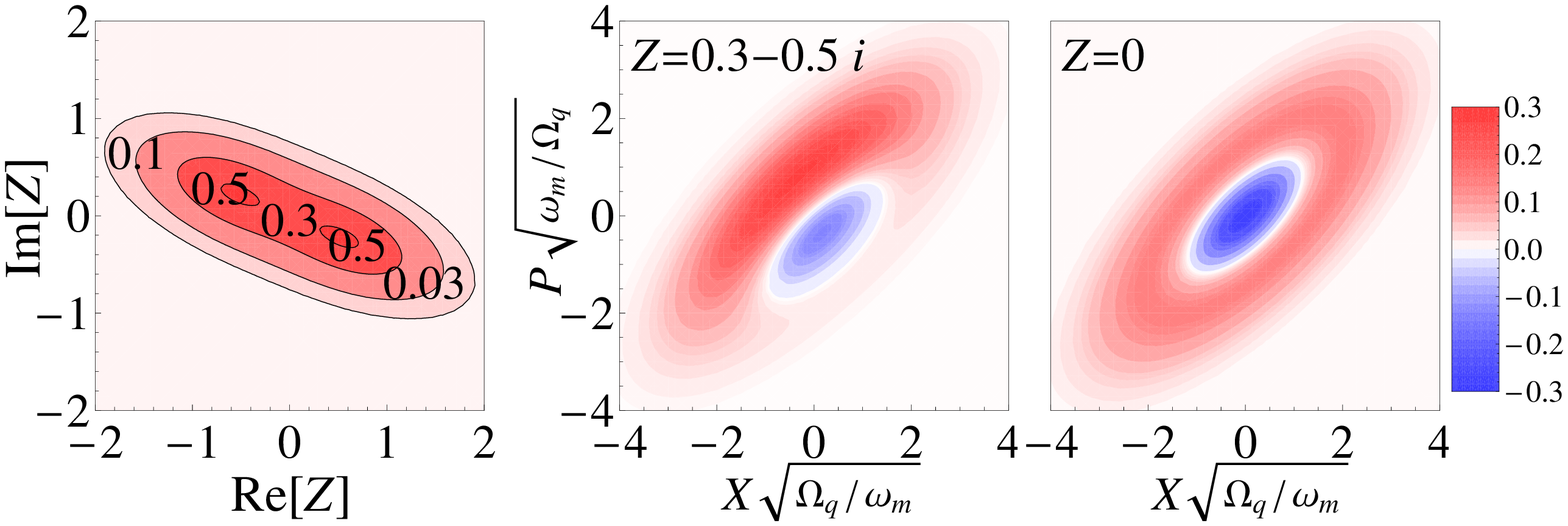}
\includegraphics[width=0.48\textwidth, bb=0 0 745 250,clip]{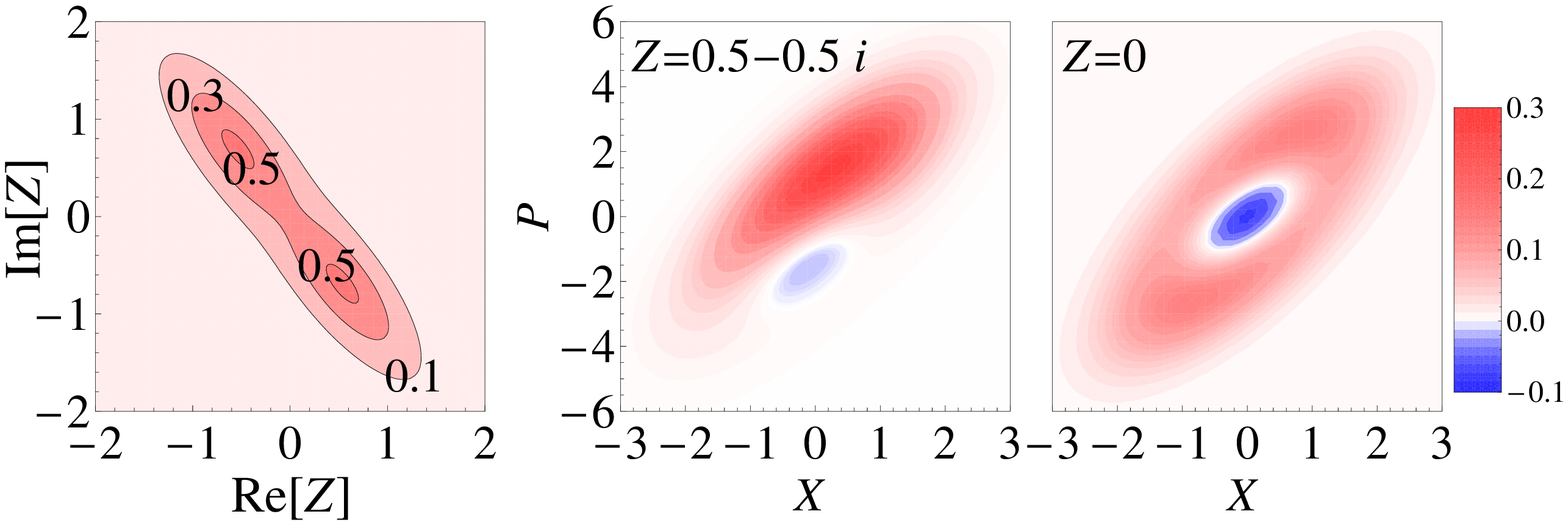}
\caption{(Color online) Distributions of measurement results (left panels) and
the corresponding Wigner function of the oscillator given the most probable
measurement results (middle panels) and less probable results but with a significant
non-Gaussianity (right panels). The upper panels show the case of non-Gaussian
state-preparation with future gravitational-wave detectors, and the lower panels for small-scale
devices. We used normalized coordinates (with respect to $x_q$ and $p_q$) and
introduced $\Omega_q\equiv \sqrt{\hbar m/\alpha^2}$. \label{Wigner}}
\end{figure}

With Eq.\,\eqref{Wig}, we can justify the simple-case qualitative results. We
use the same specifications listed in Table \ref{tab1}. As an example, we assume
a photon shape of $f(t)=\sqrt{2\gamma_f}e^{(\gamma_f+i\omega_f)t}$ and specify
that $\omega_f/2\pi=\gamma_f/2\pi=70$ Hz in the case of future gravitational-wave detectors,
and $\omega_f/\omega_m=0.1,\,\gamma_f/\omega_m=0.3$ for small-scale experiments.
The Wigner functions for some given measurement results are shown in
Fig. \ref{Wigner}. In both cases, there are negative regions in the Wigner function,
which is a unique feature of the quantumness. The prepared non-Gaussian quantum state can be
independently verified using the quantum tomography protocol developed in Ref. \cite{state_ver}
that allows sub-Heisenberg accuracy of Wigner function reconstruction, which is crucial
for revealing those negativity regions.

{\it Acknowledgment.} We thank our colleagues at Caltech Theoretical
Astrophysics group and LIGO Macroscopic-Quantum-Mechanics (MQM) group
for fruitful discussions. S.D., H.M.-E., H.Y. and Y.C. are supported by
the Alexander von Humboldt Foundation's Sofja Kovalevskaja Programme,
NSF grants PHY-0653653 and PHY-0601459, as well as the David and Barbara
Groce startup fund at Caltech. H.M. is supported by the Australian Research Council.

\end{document}